\documentclass[3p,twocolumn]{elsarticle}

\usepackage{lineno,hyperref}
\modulolinenumbers[5]

\journal{Nucl. Instru. Methods Phys. Res. B}

\begin{document}

\begin{frontmatter}

\title{Cross Sections for proton induced high energy $\gamma$-ray emission (PIGE) in reaction $^{19}$F(p,$\alpha\gamma$)$^{16}$O at incident proton energies between 1.5 and 4 MeV}

\author[myaddress,mysecondaryaddress]{P. Cabanelas\corref{correspondingauthor}}

\cortext[correspondingauthor]{Corresponding author}
\ead{pablo.cabanelas@usc.es}

\author[myaddress,otheraddress]{J. Cruz}
\author[myaddress,otheraddress]{M. Fonseca}
\author[myaddress,yetanother]{A. Henriques}
\author[myaddress]{F. Louren\c{c}o}
\author[myaddress]{H. Lu\'is}
\author[myaddress,otheraddress]{J. Machado}
\author[myaddress]{J. Pires Ribeiro}
\author[myaddress]{A. M. S\'anchez-Ben\'itez}
\author[myaddress,yetanother]{P. Teubig}
\author[myaddress,yetanother]{P. Velho}
\author[myaddress]{M. Zarza-Moreno}
\author[myaddress,yetanother,final]{D. Galaviz}
\author[myaddress,otheraddress]{A.P. Jesus}

\address[myaddress]{Nuclear Physics Center, University of Lisbon, P-1649-003 Lisbon, Portugal}
\address[mysecondaryaddress]{Particle Physics Department, University of Santiago de Compostela, Santiago de Compostela, Spain}
\address[otheraddress]{Laborat\'orio de Instrumenta\c{c}\~ao, Engenharia Biom\'edica e F\'isica da Radia\c{c}\~ao (LIBPhys-UNL), Departamento de F\'isica, Faculdade de Ci\^encias e Tecnologia, Universidade NOVA de Lisboa, P-2829-516 Caparica, Portugal}
\address[yetanother]{Laborat\'orio de Instrumenta\c{c}\~ao e F\'isica Experimental de Part\'iculas (Lip), Av. Elias Garcia 14, P-1000-149 Lisbon, Portugal}
\address[final]{Departamento de F\'isica, Faculdade de Ci\^encias da Universidade de Lisboa, Campo Grande, P-1749-016 Lisbon, Portugal}

\begin{abstract}
We have studied the high energy gamma-rays produced in the reaction $^{19}$F(p,$\alpha\gamma$)$^{16}$O for incident proton energies from 1.5 to 4.0 MeV over NaF/Ag and CaF$_2$/Ag thin targets in two different sets of data. Gamma-rays were detected with a High Purity Ge detector with an angle of 130$^{o}$ with respect to the beam axis. The cross-sections for the high energy gamma-rays of 6.129, 6.915 and 7.115 MeV have been measured for the whole group between 5 and 7.2 MeV with accuracy better than 10\%. A new energy range was covered and more points are included in the cross-sections data base expanding the existing set of data. Results are in agreement with previous measurements in similar conditions. 
\end{abstract}


\end{frontmatter}

\section{Introduction}
Particle-induced gamma-ray emission (PIGE) is a common technique used to detect and analyze elements lighter than calcium \cite{Mateus}. In particular, the analysis of monoenergetic photons from the reaction $^{19}$F(p,$\alpha\gamma$)$^{16}$O can offer applications to the analysis of fluorine concentrations, for example, in dental studies \cite{Okuyama,Komatsu}. Thus, a precise understanding of that reaction is very convenient.

\begin{figure}[htb!]
\includegraphics[width=0.48\textwidth]{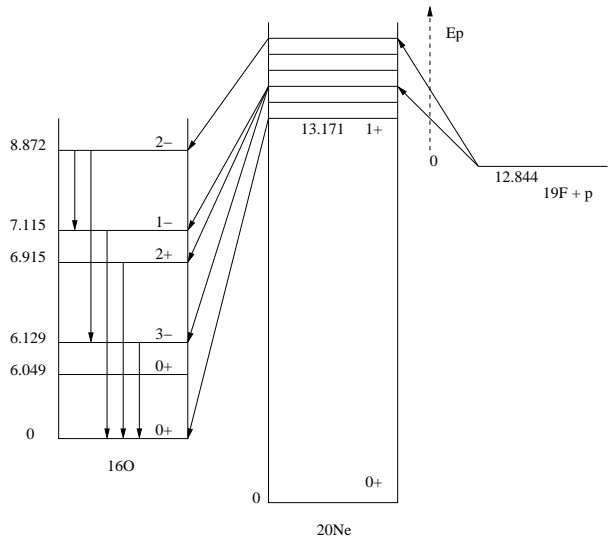}
\caption{$^{16}$O energy levels scheme. Only gamma emission transitions in excited $^{16}$O states are drawn. Transitions from $^{20}$Ne to $^{16}$O happen via $\alpha$-particle emission. All energy values are given in MeV. Drawing not to scale.} \label{fig1}
\end{figure}

The $^{19}$F(p,$\alpha\gamma$)$^{16}$O reaction has a $Q$-value of 8.11 MeV and proceeds via the $^{20}$Ne nucleus, whose $\alpha$-decay leads to the first excited states of $^{16}$O. A simplified levels scheme of $^{16}$O illustrating how the reaction proceeds mainly through the compound nucleus $^{20}$Ne is shown in Fig.~\ref{fig1}. Three of these excited states (1$^-$, 2$^+$ and 3$^-$) de-excite to the ground state (0$^+$) emitting gamma rays with energies of 6.129, 6.915 and 7.115 MeV. Besides, the fifth excited state of 8.872 MeV can be also populated and can de-excite to the 1$^-$ and 3$^-$ lower states by the emission of 1.755 and 2.741 MeV photons respectively.

The present work deals with the measurement and analysis of gamma-ray yields and cross-section of the reaction $^{19}$F(p,$\alpha\gamma$)$^{16}$O measured at a polar angle of 130$^{o}$, ranging the proton beam energy from 1.5 to 4.0 MeV. The final cross-section is calculated for the whole group of high energy gammas, from 5 to 7.2 MeV, and compared with other results from the literature in similar conditions \cite{Fessler,Caciolli,Ranken,Willard}. The cross-section values in the beam range of 1.5 and 3.0 MeV are given for the first time.

\section{Experiment}
The experimental work was carried out at the IST-CTN Tandem accelerator in Lisbon \cite{ctn_web}, in two separate stages. Proton beam energy was calibrated by the 872 keV resonance of the $^{19}$F(p,$\alpha\gamma$)$^{16}$O reaction, by the 1645 keV and 1930 keV resonances of the $^{23}$Na(p,p$'\gamma$)$^{23}$Na reaction and by the 3470 keV resonance of the $^{16}$O(p,p)$^{16}$O reaction \cite{Fonseca}. In the first stage, the measurements were performed with a proton beam at energies from 2.51 to 4.04 MeV with an approximate 4 keV step. The proton beam reached the reaction point with a 2mm diameter spot after collimators with typical currents of about 100 nA in the reaction chamber, working the last as a Faraday cup. The target consisted on a NaF thin film of 36 $\mu$g/cm$^2$ evaporated over a self-supporting Ag film of 64 $\mu$g/cm$^2$. During the second stage, a proton beam energy range from 1.50 to 2.48 MeV was covered in approximate steps of 10 keV with the same accelerator conditions. In the second stage, the target consisted on a CaF$_2$ thin film of 64 $\mu$g/cm$^2$ evaporated also over a self-supporting Ag film of 68 $\mu$g/cm$^2$. Both targets were locally developed by our group in Lisbon.

\begin{figure}[htb!]
\includegraphics[width=0.48\textwidth]{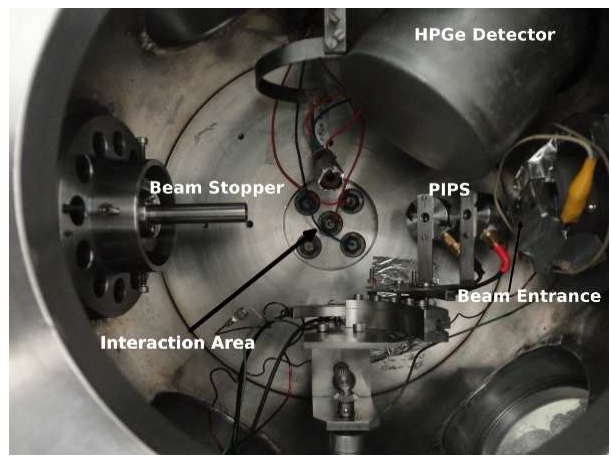}
\caption{Photograph of the experimental arrangement inside the reaction chamber. The High-Purity Ge Detector is located at an angle of 130$^{o}$ with respect to the beam line, and a distance of 55.5mm with respect to the interaction point. The PIPS detectors and their rotating system are also visible. The targets are attached to the chamber's cap.} \label{fig_chamber}
\end{figure}

The gamma radiation was detected with an ORTEC High-Purity Germanium detector (HPGe) of 64mm $\times$ 62.6mm with resolution of 1.76 keV and intrinsic relative efficiency of 45\% for the 1.33 MeV $^{60}$Co line. It was placed at an angle of 130$^{o}$ and a distance of 55.5mm with respect to the interaction point. For detecting charged particles, two Camberra Passivated Implanted Planar Silicon (PIPS) detectors of 50mm$^2$ each, effective depth of 100 $\mu$m, and 26 keV resolution for alphas of 5486 keV from $^{241}$Am, were also operative inside the reaction chamber placed at about 110$^{o}$ and 140$^{o}$ with respect to the target. Figure \ref{fig_chamber} shows the experimental arrangement inside the reaction chamber.

In the first stage of the measurements, each spectrum was acquired to the same number of total counts, resulting in about 10$^5$ counts in average in the region of 5 to 7.2 MeV, while a new current integrator was used in the second stage which allowed to record each spectrum to the same number of current collected in the reaction chamber. This method made simpler the normalization procedure for the second set of data. 

A custom made current integrator was used for the start/stop DAQ signal so that each spectrum was recorded with the same collected charge in the reaction chamber.

\section{Simulation}
The reaction chamber's HPGe detector setup was simulated in Geant4 \cite{Geant4} and data was analyzed with ROOT \cite{ROOT}, all together within an user-friendly framework for nuclear reactions simulations called EnsarRoot \cite{EnsarRoot_web}, based on the FairRoot framework \cite{FairRoot}. The detector geometry was designed to fit all the manufacturer's specifications.

\begin{figure}[htb!]
\includegraphics[width=0.49\textwidth]{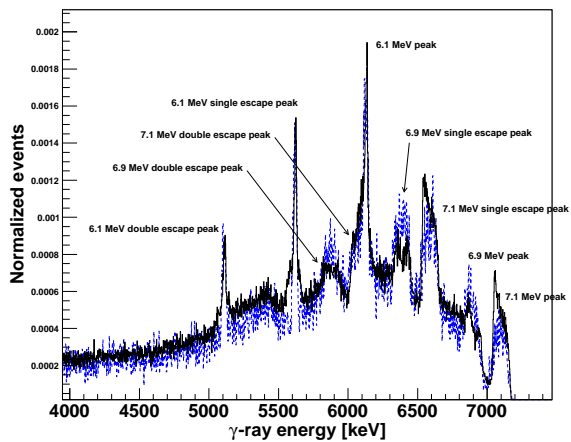}
\caption{High energy gamma-rays spectra for an incident proton beam energy of 1845 keV, both experimental (solid black line) and simulated (dashed blue line). Gamma peaks of 6.129, 6.915 and 7.115 MeV are labeled together with their corresponding single and double escape peaks.} \label{fig_specs}
\end{figure}

The corresponding spectra of the photons of interest was simulated in the HPGe detector, with the corresponding intrinsic resolution. The kinematics of the reaction was included as well in the simulation, allowing to the photons to be emitted with a random Lorentz boost ranging from $\beta=0$ to $\beta=0.02$, as they can escape with the recoil nucleus either at rest or still moving, and in any direction with respect to the beam axis. This results in the effect of having a broadening in the 6.9 MeV and 7.1 MeV photopeaks, together with their corresponding single and double escape peaks. A comparative between an experimental and a simulated spectrum can be observed in Figure \ref{fig_specs}.

\begin{figure*}[htb!]
\includegraphics[width=\textwidth]{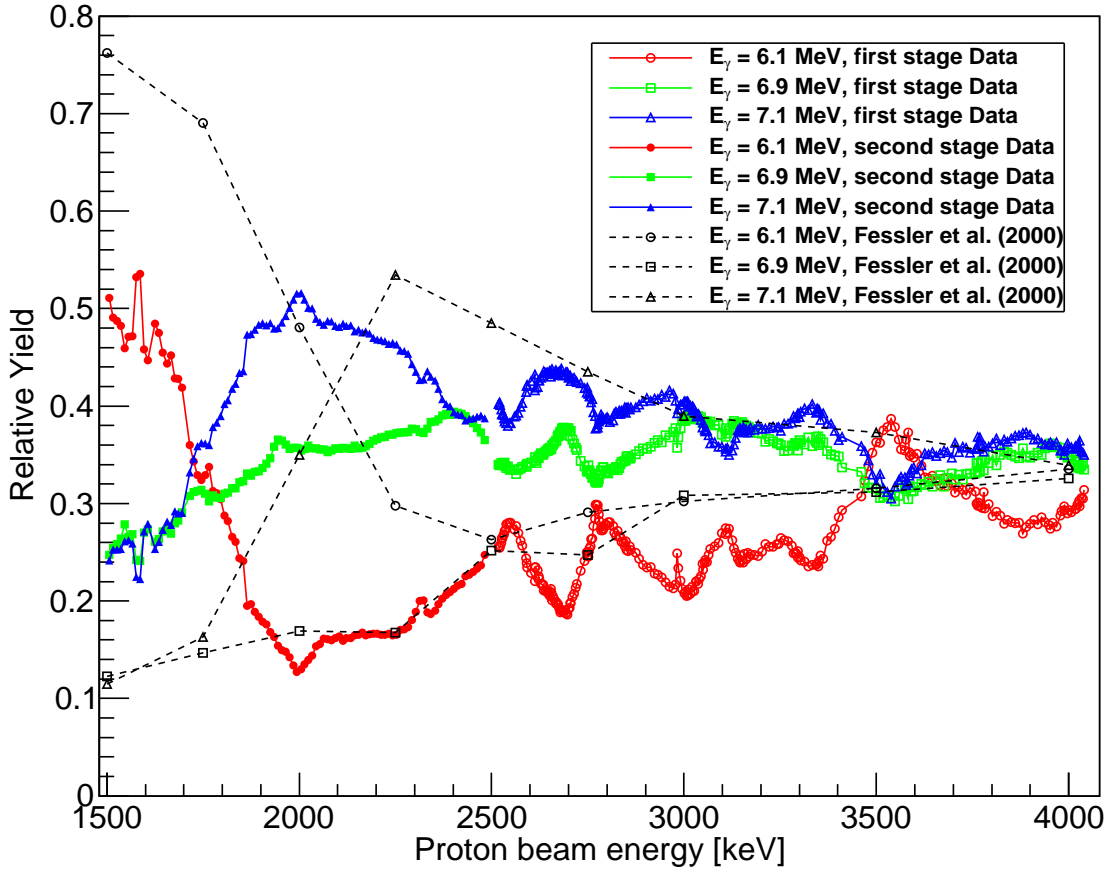}
\caption{Contribution to the total yield of the three gamma-rays of 6.1, 6.9 and 7.1 MeV, calculated by fitting the sum of the simulated spectra to the real total spectrum. Red, green and blue points represent the evolution of the 6.1, 6.9 and 7.1 MeV gammas respectively, and are compared with data from \cite{Fessler}, represented with open markers and black dashed line.} \label{fig_yields}
\end{figure*}

It is known that the contribution of each peak to the total spectrum changes as the proton beam energy changes \cite{Fessler}. This was included in the simulation by giving individual weights to the sum of the three simulated transitions spectra and fitting that weights to the corresponding real spectrum by a Least Square Method. Results of the individual contributions are shown in Figure \ref{fig_yields}. The comparison with data from \cite{Fessler} is also shown.

The main contribution of the simulation to this analysis is a much better estimation and understanding of the detector efficiency at higher energies. The detection efficiency was calculated in simulation for the photon energy range between 4.5 and 7.5 MeV, and a parameterization of the detector efficiency as a function of the energy of the registered photon, $\varepsilon_{\gamma}(E_{\gamma})$, was obtained. The efficiency was then compared with that at high energies extrapolated from measurements with radioactive sources resulting in a good agreement. Our region of interest (5 MeV to 7.2 MeV) falls in the tail of the efficiency curve \cite{Rodenas}, but the value over the region is far to be constant. Furthermore, a difference of 20\% is observed between both ends of the range. By assuming a constant efficiency in the range like the average value, as it is done in \cite{Caciolli}, the final emission yield of a given photopeak, and therefore the cross-section, can be underestimated by a factor up to 5\%. In this work, the parameterization $\varepsilon_{\gamma}(E_{\gamma})$ was applied bin by bin to each acquired spectrum to obtain the yield and the resulting contribution was used for the cross section calculation.

\section{Results and Data analysis}

A typical spectrum of the studied gamma-ray energy range is shown in Figure \ref{fig_specs}; in this case, the acquired spectrum for proton beam energy of 1845 keV is shown (solid black line). Also, the simulated spectrum (dashed blue line) obtained taking into account the contribution of each peak at the given energy is shown. The three main peaks are identified together with their single and double escape peaks.

The differential cross-section for the high energy gamma-rays of 6.129, 6.915 and 7.115 MeV was obtained for the whole group, within a window between 5 and 7.2 MeV in the energy spectrum of gammas. The cross section values were deduced using: 

\begin{equation} \label{eq1}
\frac{d\sigma}{d\Omega} = \frac{N^{F}_{\gamma}}{N^{Ag}_{p}} \cdot r \cdot \frac{\varepsilon_p}{\varepsilon_{\gamma}(E_{\gamma})} \cdot \left(\frac{d\sigma}{d\Omega}\right)_{Ruth}^{Ag}
\end{equation}

where $N^{F}_{\gamma}$ is the counting for the high-energy gammas in the given window; $N^{Ag}_{p}$ is the area of the proton backscattering peak on silver; $r$ is a stoichiometric factor which represents the relation between the density of atoms of silver and fluorine in the target ($r=\frac{N_{Atms,Ag}}{N_{Atms,F}}$), and has been obtained by alpha particle RBS analysis of the sample; $\varepsilon_p$ is the absolute efficiency of the particle detector for protons; $\varepsilon_{\gamma}(E_{\gamma})$ is the parameterized efficiency for gammas in the energy window; $\left(\frac{d\sigma}{d\Omega}\right)_{Ruth}^{Ag}$ is the proton Rutherford backscattering cross-section on silver, which can be calculated analytically.

\begin{figure*}[htb!]
\includegraphics[width=\textwidth]{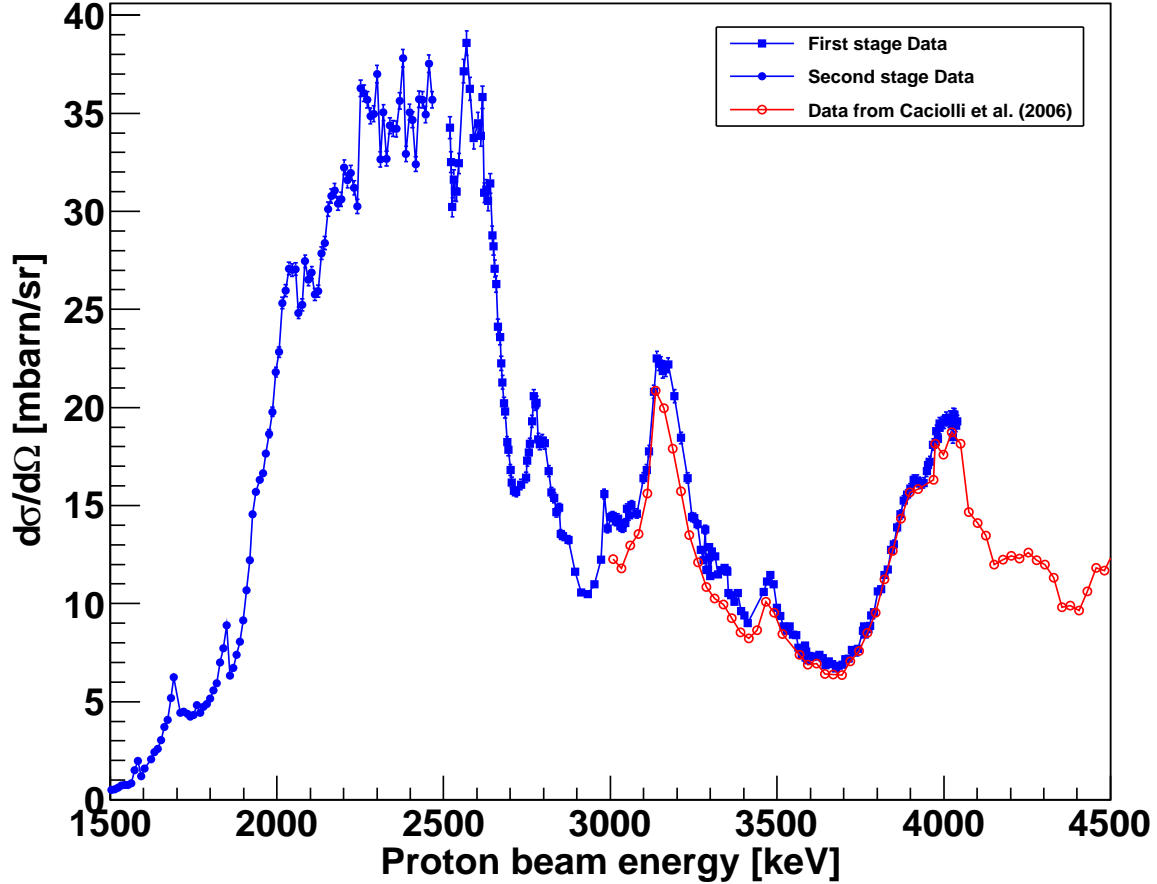}
\caption{Differential Cross-Sections for the whole group of high energy $\gamma-rays$ in the range of 5 and 7.2 MeV. Both sets of analyzed data in this paper are shown, together with data from \cite{Caciolli}} \label{fig_cs_2}
\end{figure*}

The measured differential cross-sections for the whole group are presented in Fig. \ref{fig_cs_2}. The contributions to the error bars come from statistical errors, resulting in any case lower than 10\%. Systematic errors coming from the Rutherford cross-section calculation of Ag, stoichiometric factor calculation and proton absolute efficiency are estimated to be less than 2\%. On the other hand, it is known that the beam line can produce a thin deposit of carbon over the target \cite{Fonseca}. Thus, the effective beam energy was calculated by taking into account the proton energy loss in the carbon. The Rutherford Back-Scattering technique (RBS) with data from the PIPS was applied to estimate the amount of carbon in the target for each single run. Due to technical issues, this correction could only be applied to the second stage set of data.  

Results from both sets of measurements are compared to those from \cite{Caciolli}. As it can be observed, a large portion of the data points from the second stage set appears to be shifted by about 10 keV with respect to \cite{Caciolli}. This can be attributed to the fact that the beam energy value is not corrected by energy losses in the target, as mentioned above. Overall the agreement in shape is very good although our cross section is slightly higher (between 3 and 10\%) in part of the overlapped range.

Values for cross-sections are not given in Refs. \cite{Willard} and \cite{Fessler}, but yields. Our data also agrees with them both in shape and in resonances, although due to a finer step in proton beam energy more resonances can be found in our case. Besides, our measuring angle is not the same than that for \cite{Willard}, \cite{Ranken} or \cite{Fessler}, and not even than that for \cite{Caciolli}. Although \cite{Fessler} declares isotropic gamma emission at proton energies of 2.0 and 3.0 MeV within 15\%, and the same assumption is done in \cite{Ranken}, we believe that angular distribution of photon emission is not flat, and also changes with proton beam energy, since the contribution of each peak also depends on the proton beam energy.   

Cross-sections values in tabular form for their use in PIGE analysis are available upon request.

\section{Conclusions}
The high energy gamma-ray emission of the $^{16}$O first excited states was studied with the reaction $^{19}$F(p,$\alpha\gamma$)$^{16}$O at beam energies from 1.5 to 4.0 MeV. The corresponding gamma spectrum was studied and reproduced in simulation, and the contribution of each gamma-ray to the total spectrum was calculated. The detection efficiency was calculated as well and characterized in the range of interest.

The differential cross-section for the whole group of gamma rays between 5 and 7.2 MeV was calculated and the obtained curves are in agreement with previous measurements in similar conditions. The energy range for the available cross-section data has been increased for PIGE requirements. 

\section*{Acknowledgements}
This work has been financially supported by Plan I2C 2013 by Xunta de Galicia (ES), and by CFNUL strategic project PEstOE/Fis/UI0275/2014.

The authors wish to thank their CTN collaborators and technical staff for their support during the measurements. 

\section*{References}

\bibliography{mybibfile}

\end{document}